\documentclass[12pt,a4paper]{article}
\usepackage{color,graphics,amsmath,epsfig,rotating}

\usepackage{amsmath}
\usepackage{amssymb}

\usepackage{graphicx}

\usepackage{cancel} 
\begin{document}

\newcommand{\ds}{\texttt{DarkSusy}}
\newcommand{\micro}{\texttt{micrOMEGAs}}
\newcommand{\py}{\texttt{PYTHIA}}
\newcommand{\HE}{\texttt{HERWIG}}
\newcommand{\msun}{{\rm M_{\odot}}}
\newcommand{\ie}{{\it i.e.} }
\newcommand{\eg}{{\it e.g.} }
\newcommand{\pbar}{${\rm \bar{p}}$}
\newcommand{\dbar}{${\rm \bar{D}}$}
\newcommand{\beq}{\begin{equation}}
\newcommand{\eeq}{\end{equation}}
\def\rsun{r_\odot}
\newcommand{\noi}{\noindent}

\newcommand{\calchep}{\texttt{CalcHEP}}
 \def\lnTop{\log\left(\frac{m_t^2}{Q^2}\right)}
\def\lsp{\chi}
\def\mlsp{m_{\chi}}
\def\mglu{m_{\tilde g}}

\begin{center}

{\Large\bf micrOMEGAs4.1: two dark matter candidates} \\[8mm]

{\large   G.~B\'elanger$^1$, F.~Boudjema$^1$, 
A.~Pukhov$^2$, A.~Semenov$^3$ }\\[4mm]

{\it 1) LAPTH, Univ. de Savoie, CNRS, B.P.110,  F-74941 Annecy-le-Vieux, France\\
    2) Skobeltsyn Inst. of Nuclear Physics, Moscow State Univ., Moscow 119992, Russia\\
      3) Joint Institute of Nuclear research, JINR, 141980 Dubna, Russia  }\\[4mm]

\end{center}

\begin{abstract}
\micro~ is a code to compute dark matter observables  in generic extensions of the standard model.
This version of \micro~ includes a generalization of the Boltzmann equations to take into account the possibility of two dark matter candidates. 
The modification of the relic density calculation to include interactions between the two DM sectors as well as semi-annihilation  is presented. Both DM signals in direct and indirect detection are computed as well. An extension of the standard model with two scalar doublets and a singlet is used as an example.
\end{abstract}

\section{Introduction}

Strong evidence for dark matter at the scale of galaxies and galaxy clusters is sustained by recent precise cosmological observations by the PLANCK satellite~\cite{Ade:2013lta}. However,
the simplest WIMP paradigm, e.g. within the framework of SUSY, is challenged by collider data since  no evidence for new particles was found in the first LHC run~\cite{ATLAS_NP,CMS_pMSSM}. At the same time several anomalies have been observed in both direct detection ~\cite{Bernabei:2010mq,Aalseth:2010vx,Aalseth:2012if,Angloher:2011uu,Agnese:2013rvf} and indirect detection experiments~\cite{Weniger:2012tx,FermiLAT:2011ab,Adriani:2008zr,Aguilar:2013qda}. 
The various signals corresponding to vastly different mass scales cannot be explained by a single dark matter candidate. 
Moreover some of these anomalies hint at cross sections stronger than the canonical value deduced from cosmological observations.
While these anomalies cannot be unambiguously  associated with dark matter, for example pulsars could be the source of the higher than expected positron flux at high energies~\cite{Hooper:2008kg}, and anomalies in direct detection corresponding to light dark matter are challenged by competing experiments with null results ~\cite{Aprile:2012nq,Akerib:2013tjd}, these observations 
raise the interesting possibility that the anomalies could be due to two dark matter candidates.

On the theoretical side, multi-component DM models  have been considered a long time ago.  For example the idea that the neutrino, the axion or its supersymmetric partner, the axino could constitute  a fraction of the total DM has been examined carefully over the years~\cite{Hannestad:2003ye,Hannestad:2010yi,Baer:2009ms,Bae:2013hma}. Models where both components are WIMPs - and could therefore lead to typical signatures at different mass scales - have also been examined
~\cite{Ma:2006km,Zurek:2008qg,Batell:2010bp,Fukuoka:2010kx,Belanger:2012vp,Aoki:2012ub,Ivanov:2012hc,Chialva:2012rq,Heeck:2012bz,Modak:2013jya,Aoki:2013gzs,Geng:2013nda,Kajiyama:2013rla,Bhattacharya:2013hva,Belanger:2014bga}.

When DM is made of two WIMPs  the interactions between the two dark matter - so-called dark matter conversion~\cite{Liu:2011aa,Belanger:2011ww}, modify the Boltzmann equation and impact the computation of the relic density. Here 
we generalize the \micro~ routine to compute the relic density to include  all possible interactions between the particles in the dark sectors. In particular all  semi-annihilation processes (where two dark matter particles annihilate into another dark matter particle and a standard one) are also included. We also add a few facilities to the direct and indirect detection routines to take into account the contribution of each component to the DM density. 
As a working example of  a multi-component DM model we consider an extension of  the SM containing one extra scalar  doublet and one scalar singlet. A $Z_4$ discrete symmetry leads to two stable DM candidates and implies both semi-annihilations and self-interactions between the two dark sectors~\cite{Belanger:2012vp,Belanger:2014bga}.

This paper is organized as follows. In section 2 we list all possible DM interactions and mention the  $Z_N$ or $Z_N\times Z_M$ discrete groups that could lead to these interactions.   The generalization of the relic density calculation and the method used to solve these equations are described in section 3. The modification of the micrOMEGAs routines that deal with two DM candidates are described in section 4.
Section 5 contains sample results obtained for the $Z_4$ model with two doublets and a singlet.

\section{Classification of 2-DM models.}

 \micro~ exploits the fact that models of dark matter exhibit a discrete symmetry
and that the fields  of the model transform as 
\begin{equation}
   \phi \to e^{i2\pi X_{\phi}} \phi
\end{equation}
where the charge $|X_{\phi}|<1$. 

The particles of the  Standard Model 
are assumed to transform trivially under the discrete symmetry, $X_\phi=0$. The 
lightest particle with charge $X_\phi\neq 0$  will be  stable and, if neutral, can be considered as a DM candidate.
Typical  examples  of discrete symmetries used for constructing single DM models  are $Z_2$ and  $Z_3$. 
Multi-component DM can arise in models with larger discrete symmetries. A simple 
example is a model with  $Z_2\times Z_2'$ symmetry, the particles charged under  $Z_2$($Z_2'$) will belong to the first (second) dark sector. The lightest particle of each sector - whether it is a fermion or a scalar - will be stable and therefore a potential DM candidate. 
Another example is a model with a $Z_4$  symmetry.  The two dark sectors contain particles with $X_\phi=\pm 1/4$ and $X_\phi=1/2$ respectively. The lightest particle with  charge $1/4$ is always stable while the lightest particle of charge $1/2$ is stable only if its decay into two particles of charge $1/4$ is kinematically forbidden,  

We can write all possible dimension-4  interactions for models with two dark sectors. 
Let $\phi_a$ and $\phi_b$ be the generic names of particles belonging to each dark sector, with $X_a\neq \pm X_b\neq 0$.
For any choice of discrete group, the potential  for scalars can contain the  terms 
\footnote{We do not write the generic  Lagrangians when the dark sector contains fermions and gauge bosons, this Lagrangian is simpler than for scalars and can be easily adapted for each choice of symmetry and particle content.}
\begin{equation}
\label{common}
 V_0 =   m_a^2 \phi_a\bar\phi_a + m_b^2 \phi_b\bar\phi_b + \lambda_a \phi_a^2\bar\phi_a^2  + \lambda_b  \phi_b^2\bar\phi_b^2 + \lambda_{ab} \phi_a\bar\phi_a \phi_b\bar\phi_b\end{equation}
Here we omit isospin indices and assume that $\phi_a(\phi_b)$ represent all the different scalar particles 
with a given discrete charge. Additional terms are possible depending on the choice of the symmetry group.  The list of all  possible structures for models with two scalar dark sectors is given in Table \ref{ListOfGroups}  
together with the lowest $Z_N$ or $Z_N\times Z_M$ symmetry that leads to such interactions. When the dark sector contains fermions and gauge bosons, the number of allowed terms is more limited and can be easily written for each specific case. 
\begin{table}[]
 \label{ListOfGroups}
\begin{tabular} {|l|l|l|l|l|}
\hline
title & terms   & group & $X_a$ & $X_b$\\  
\hline
&&&&\\
L2x2 &$ \phi_a^2 + \phi_a^4 +\phi_b^2 + \phi_b^4 + \phi_a^2 \phi_b^2 +\bar\phi_a^2 \phi_b^2  $ &$Z_2\times Z_2  $ &$   k/2     $ &$     n/2         $  \\
L3x3 &$ \phi_a^3 + \phi_b^3                                                $ &$Z_3\times Z_3  $ &$   k/3     $ &$      n/3         $   \\
L2x3 &$ \phi_a^2+\phi_a^4 + \phi_b^3                                            $ &$Z_2\times Z_3  $ &$   k/2     $ &$     n/3         $   \\
L4   &$ \phi_a^2 \phi_b + \phi_a^4 + \phi_b^2 + \phi_b^4 + \phi_a^2\bar\phi_b          $ &$Z_4            $ &$   k/4     $ &$    -k/2        $   \\      
L5   &$ \phi_a^3 \phi_b + \phi_a^2\bar\phi_b +\phi_b^3 \bar\phi_a +\phi_a\phi_b^2$ &$Z_5            $ &$   k/5     $ &$     -3k/5       $   \\
L6a  &$ \phi_a^2 \phi_b + \phi_b^3 + \bar\phi_a^2\phi_b^2                    $ &$Z_6            $ &$    k/3+n/2$ &$    k/3    $   \\    
L6b  &$ \phi_a^3 \phi_b + \phi_b^2 +\phi_b^4 +\phi_a^3\bar\phi_b                   $ &$Z_6            $ &$   k/6     $ &$      -k/2$   \\
L7   &$ \phi_a^2 \phi_b +\bar\phi_a \phi_b^3                            $ &$Z_7            $ &$     3k/7  $ &$      k/7  $   \\  
L8a  &$ \phi_a^2 \phi_b + \phi_b^4                                         $ &$Z_8            $ &$    -k/8   $ &$     k/4    $   \\
L8b  &$ \phi_a \phi_b^3 + \phi_a^3 \phi_b +\bar\phi_a^2 \phi_b^2              $ &$Z_8            $ &$   -3k/8   $ &$     k/8     $   \\
L9   &$ \phi_a^3 \phi_b + \phi_b^3                                         $ &$Z_9            $ &$   k/9     $ &$     -2k/3  $   \\
L12  &$ \phi_a^3 \phi_b + \phi_b^4                                         $ &$Z_{12}         $ &$   k/12    $ &$     -k/4    $   \\
\hline
\end{tabular}
 \caption{ Table of generic group structures and interaction vertices  for
2-component DM models, here $k,n$ are integers. Hermitian conjugated terms are omitted,  the complete list can be obtained 
swapping    $ \phi_a \leftrightarrow \bar\phi_a $ and $\phi_a \leftrightarrow \phi_b$ .}
\end{table}     

The Lagrangian for a concrete multi-component DM model will of course depend on the spin and isospin of the non-standard particles.
Consider a model with two scalar doublets $H_1$  and $H_2$ and a singlet, $S$. We impose a $Z_4$ symmetry with $X_{H_1}=0$, $X_{H_2}=1/2$ and $X_{S}=1/4$.  Both $H_2$ and $S$ are inert, i.e. they do not couple to fermions while $H_1$ has couplings similar to those of the SM Higgs, for more details see Ref.~\cite{Belanger:2012vp}.
The field $\phi_a$ stands for the singlet and $\phi_b$ with the doublet. The potential reads

\begin{equation}
\begin{split}   
V_{Z_{4}} &= V_{0} + \frac{\lambda_{S}}{2} (S^{4} + S^{\dagger 4}) +\frac{\lambda_{5}}{2} \left[(H_{1}^{\dagger} H_{2})^{2} + (H_{2}^{\dagger} H_{1})^{2} \right] \\
          &+ \frac{\lambda_{S12}}{2} (S^{2} H_{1}^{\dagger} H_{2} + S^{\dagger 2}H_{2}^{\dagger} H_{1}) +\frac{\lambda_{S21}}{2} (S^{2} H_{2}^{\dagger} H_{1} + S^{\dagger 2} H_{1}^{\dagger}H_{2}),\\
\end{split}
\label{eq:pot:Z4}
\end{equation}

where 

\begin{equation}
\begin{split}   
  V_{0}&=  \mu_{1}^{2} |{H_{1}}^{2}| + \lambda_{1} |{H_{1}}^{4}|
  + \mu_{2}^{2} |H_{2}|^{2} + \lambda_{2} |H_{2}|^{4} + \mu_{S}^{2} |S|^{2}
+ \lambda_{S} |S|^{4} \\
  &+ \lambda_{S1} |S|^{2} |H_{1}|^{2}
  + \lambda_{S2} |S|^{2} |H_{2}|^{2} + \lambda_{3} |H_{1}|^{2} |H_{2}|^{2}
  + \lambda_{4} (H_{1}^{\dagger} H_{2}) (H_{2}^{\dagger} H_{1}).
\end{split}
\label{eq:Vc}
\end{equation}.

\section{Relic density computation}
\label{dmRelic}

\subsection{Evolution equations}

The derivation of  the equations for the number density of two DM particles is based on  standard assumptions: 
1) all particles of the same sector are in thermal equilibrium 
b) particles of the dark sectors have the same kinetic temperature as those of the SM 
c)  the number densities  of DM particles can differ from the equilibrium values  when  the number density of  DM  particles times their annihilation cross section  becomes too low to keep up with the expansion rate of the Universe.  The different processes that influence the number densities of DM include 
the annihilation and co-annihilation processes in each sector $\phi_\alpha\phi_\alpha^*\rightarrow X X$, the DM conversion processes $\phi_\alpha\phi_\beta\rightarrow \phi_\gamma\phi_\delta$, and the semi-annihilation processes
$\phi_\alpha\phi_\beta\rightarrow\phi_\gamma X$. Here X denotes any SM particle and $\phi_\alpha$ stands for any particle in the dark sectors. 
The equation for the number densities, $n_a$  of DM particles in sector 1 and 2  reads

\begin{eqnarray}
\label{t-evolution}
\frac{dn_a}{dt}&=&-\sigma_v^{aa00}  \left(n_a^2-\bar{n}_a^2 \right) -
\sigma_v^{aab0}\left( n_a^2- n_b \frac{\bar{n}_a^2}{\bar{n}_b} \right)
- \sigma_v^{aabb}\left( n_a^2- n_b^2 \frac{\bar{n}_a^2}{\bar{n}_b^2}
\right)\nonumber  \\
  && - \frac{1}{2}\sigma_v^{aaa0}  \left(n_a^2-\bar{n}_a n_a \right)
     - \frac{1}{2}\sigma_v^{aaab}\left( n_a^2- n_a n_b \frac{\bar{n}_a}{\bar{n}_b}\right)
      - \frac{1}{2}\sigma_v^{abbb}\left( n_a n_b- n_b^2\frac{\bar{n}_a}{\bar{n}_b}\right)\nonumber \\
               &&  -\frac{1}{2}\sigma_v^{abb0}\left( n_a n_b- n_b \bar{n}_a \right)
+\frac{1}{2}\sigma_v^{bba0}(n_b^2-n_a\frac{\bar{n}_b^2}{\bar{n}_a})   - 3H n_a 
\end{eqnarray}
Here $b$ denotes DM sector different from $a$ ($a\ne b$), $\bar{n}_a$ and
$\bar{n}_b$    denote the equilibrium number densities of particles in the two  dark sectors. 
The label $0$ is used for SM particles,
$\sigma_v^{abcd}$  means  the thermally averaged cross section defined as
\begin{eqnarray}
\sigma_v^{abcd}(T) &=& \frac{T}{8\pi^4  \overline{n}_a(T)\overline{n}_b(T)}\int ds\sqrt{s} K_1\left(\frac{\sqrt{s}}{T}\right)   
\sum_{\substack{
\alpha\in a\; \beta\in b\;\\ \gamma\in c\; \delta\in d \\\mathrm{pol.}}}  
p_{\alpha\beta }^2(s) {\sigma}_{\alpha\beta\to \gamma\delta}(s),
\label{sigmaV}\\
\overline{n}_a(T)&=&\frac{T}{2\pi^2 } \sum_{\alpha\in a} g_\alpha m^2_\alpha K_2(\frac{m_\alpha}{T}),
\end{eqnarray}
Here ${\sigma}_{\alpha\beta\to \gamma\delta}$ is the cross section for  the process $\phi_\alpha \phi_\beta\rightarrow \phi_\gamma \phi_\delta$,  $K_1,K_2$ are modified Bessel functions of the second kind, and 
$m_\alpha$ and $g_\alpha$ stand for the  mass and the number of degrees of freedom of particle $\phi_\alpha$.
Roman indices take the value $0,1,2$ and  Greek indices are used to designate particles in a given sector.  The inverse reactions  are related via the detailed balance equation 
\begin{equation}
       \bar{n}_a \bar{n}_b \sigma_v^{abcd} = \bar{n}_c \bar{n}_d \sigma_v^{cdab}.
\end{equation}
Note that in a particular model, only a subset of all possible $2 \to 2 $ processes for DM annihilation listed in Table \ref{ListOfGroups} will be allowed,
and only the relevant terms will be included in Eq.~\ref{t-evolution}  by {\tt micrOMEGAs}.\footnote{Note that 3-body final states from virtual W and Z exchange are also included in the annihilation (coannihilation) processes entering the thermally averaged cross section by setting  the switches VWdecay,VZdecay =1(2). }.

Usually the DM evolution equations are solved in terms of the abundance, $Y_a=n_a/s$, where $s$ is the entropy density. 
The equation for entropy conservation 
\begin{equation}
   \frac{ds}{dt}=-3Hs
\end{equation}
allows to convert the time evolution equation into an evolution with respect to the entropy density. 
Introducing $\Delta Y_a = Y_a - \overline{Y}_a = \frac{n_a-\bar{n}_a}{s} $, Eq.~(\ref{t-evolution}) takes the simple form 
\begin{equation}
\label{MatrixEq}
        3H\frac{d\Delta Y_a}{ds} = -C_a +A_{ab}(s)\Delta Y_b +Q_{abc}(s) \Delta Y_b \Delta Y_c,
\end{equation}
where

\begin{eqnarray}
 C_a &=& 3H\frac{d \overline{Y}_a}{ds},\\
A_{aa}&=& \overline{Y}_a( 2( \sigma_v^{aa00} + \sigma_v^{aab0} + 
\sigma_v^{aabb}) + \frac{1}{2}(\sigma_v^{aaa0} +\sigma_v^{aaab}))\nonumber\\
   &+& \frac{1}{2}\overline{Y}_b(\sigma_v^{abb0}+ \sigma_v^{abbb} +
       \frac{\overline{Y}_b}{\overline{Y}_a}\sigma_v^{bba0}  )\\ 
A_{ab}&=& -\overline{Y}_a( \frac{1}{2}\sigma_v^{abaa}+\frac{1}{2}\sigma_v^{abbb}
+\overline{Y}_a/\overline{Y}_b \sigma_v^{aab0} ) - \overline{Y}_b(
   2\sigma_v^{bbaa}  +\sigma_v^{bba0} )\\
Q_{aaa}&=&   \sigma_v^{aa00} + \sigma_v^{aab0} +
\sigma_v^{aabb} + \frac{1}{2}(\sigma_v^{aaa0} +\sigma_v^{aaab})\\
Q_{aab} &=&  \frac{1}{2}(\sigma_v^{abbb}+\sigma_v^{abb0}- \sigma_v^{abaa})  \\
Q_{aba}&=&0\\
Q_{abb}&=&  -\sigma_v^{bbaa} -\frac{1}{2}\sigma_v^{bbab} -\frac{1}{2}\sigma_v^{bba0}
\end{eqnarray}  
Here as above  $b\ne a$. Since $s$ is a known function of temperature T, Eq.~(\ref{MatrixEq}) is actually the temperature  evolution equation. 

\subsection{Solution of equations}

At  temperatures larger  than the masses of DM particles  $\overline{Y}_a$ is constant and represents the fraction of total degrees of freedom for each  dark sector. 
Since  it is  constant, $C_a=0$, and the solution of Eq.(\ref{MatrixEq}) is $\Delta Y_a =0$, which means that the DM particles are in thermal equilibrium with SM particles. Note that we make the approximation that $\Delta Y \le \bar{Y}$  such that terms in $\Delta Y_j \Delta Y_k$ in Eq.(\ref{MatrixEq}) are neglected.  
Small deviations from equilibrium  are obtained by solving
\begin{equation}
    \label{largeTz4}
     \Delta Y(s)=  A^{-1}(s) C(s).
\end{equation}

The numerical  solution of Eq.~\ref{MatrixEq} used in the case of one component DM needs to be adapted, the problem is caused by a very small step size in the integration routine. In the  region where the linear term dominates,  the step of the integration over $\log(s)$  is about $H/(sA)$ leading to a very small step size when $T \approx M_{cdm}$. To bypass this problem in
the  one component DM case, we used the approximation (\ref{largeTz4}) until
\begin{equation}
\label{start}
 \Delta Y_i \approx 10^{-2}\overline{Y}
\end{equation}
At this point we switch to the numerical solution  using  the standard Runge-Kutta method. 
For two components DM, the matrix $A_{ij}$ has  two eigenstates.  When there is a noticeable  mass difference between the two  DM particles, the 
freeze-out of the heavy component occurs much before that of the light one. Thus we have a region  of temperatures where  the
approximation (\ref{largeTz4}) does not apply because the eigenvalue of one of the eigenstate  of $A$  is small, however the direct numerical integration stalls because the large  eigenvalue forces a very small step of integration. In fact
it means that  the space of solutions  has an attractor line. Equations which pose such numerical problems are known as  {\it stiff} equations. To solve such equations the {\it backward } scheme is used. In this scheme  at each step of integration one evaluates derivatives at the final point of the step rather than at the initial point  as in the standard scheme. In \micro~ we use the 
Rosenbrock algorithm \cite{Press92numericalrecipes,hairer2010solving} for solving  stiff equations. This method finds a solution for points where the standard Runge-Kutta method fails. 
In the current version we use the Fortran code presented in \cite{hairer2010solving}.

To speed up the calculation we first tabulate  different cross sections as a function of the temperature  in the interval  
$T\in[\rm{Tend},\rm{Tstart}]$, where  $\rm{Tstart}$ is the temperature for which the condition Eq.~\ref{start} is satisfied 
after solving Eq.~\ref{largeTz4}, and $\rm{Tend}=10^{-3}GeV$.   Functions which interpolates the tabulated  data are accessible to the user after the calculation of the relic density. These functions have the generic name
{\tt vs$ijkl$F(T)} where $i,j,k,l$ can take the value {\it  0,1,2} and 
$0< i \le j$, $k \ge l$. 
The  temperature dependence  of the equilibrium abundances can also be called by the user, the functions are named {\tt Y\_1(T)} and {\tt Y\_2(T)} and are defined only in the  interval $T\in
[\rm{Tend},\rm{Tstart}]$.

\section{Two DM models in micrOMEGAs}

In previous versions of \micro~\cite{Belanger:2004yn,Belanger:2006is} we assumed that the names of all particles transforming non-trivially under  the discrete symmetry 
group started with '\verb|~|'.   In the current version we need to distinguish the particles with different transformation properties with 
respect to the discrete group, that is particles belonging to different 'dark' sectors.
For this we use the convention that the names of particles in the second 'dark' sector starts with '\verb|~~|'. 
Note that \micro~ does not check the symmetry of the Lagrangian, it assumes that the name convention 
correctly identifies  all particles with the same discrete symmetry quantum numbers.

Before  evaluating  DM observables in \micro~ one needs to  call the initialization routine\\
 \noindent
 $\bullet$ \verb|sortOddParticles(name)|\\
which fills the  global parameters  presented in Table~\ref{TwoDm}. 
 \begin{table}[htbp]
 \caption{Evaluated global  variables}
 \label{TwoDm}
\begin{center}
\begin{tabular}{|l|l|l|l|}
\hline
  Name      & units          & comments                                \\  \hline
  CDM1      &{\it character} & name of the lightest particle in first DM sector \\
  CDM2      &{\it character} & name of the lightest particle in second  DM sector\\
  Mcdm1     &  GeV           & Mass of CDM1   \\
  Mcdm2     &  GeV           & Mass of CDM2        \\
  Mcdm      &  GeV           & $\min(Mcdm1,Mcdm2)$ if both exist   \\
\hline
\end{tabular}
\end{center}
\end{table}
Note that there is no restriction on the relative values of  {\tt Mcdm1}  and  {\tt Mcdm2}, either can be the lightest one. 
This  micromegas4.X version also works for models with only one DM candidate. 
In this case  {\tt CDM1} or {\tt CDM2} (depending on the name convention chosen by the user) will be initialized by NULL in C and  a blank
string in Fortran. The corresponding mass will be set to zero.  
The return parameter {\it name} contains the name of the lightest particle 
and {\tt Mcdm} its mass. If \micro~ gets NAN while evaluating constraints, then {\it name}  contains  
the name  of the problematic constraint and {\tt sortOddParticles} returns an error code.

There are two functions for the evaluation of  the relic density. The new routine\\ 
$\bullet$\verb|darkOmega2(fast, Beps)|\\
calculates $\Omega h^2$ for both one- and  two-components DM models.
The parameter {\tt fast=1} flag forces the fast calculation (for more details see
Ref.~\cite{Belanger:2004yn}). This is the recommended option and
gives an accuracy around $1\%$. The parameter {\tt Beps} defines the
criteria for including a given coannihilation channel in the computation of the
thermally averaged cross-section,~\cite{Belanger:2004yn}.  The
recommended value is {\tt Beps}=$10^{-4} - 10^{-6}$, if {\tt Beps}$=1$ only annihilation of the
lightest odd particle is computed.
 \verb|darkOmega2| also calculates the  global parameter {\tt fracCDM2} which represents the mass fraction of CDM2 
in the total relic density
\begin{eqnarray}
  \Omega &=& \Omega_1+\Omega_2\\
  \verb|fracCDM2|&=&\frac{\Omega_2}{\Omega}
\end{eqnarray}
This parameter is then used in routines which calculate the total signal from both  DM candidates in direct, indirect and neutrino telescope experiments,
 \verb|nucleusRecoil|, \verb|calcSpectrum|,  and \verb|neutrinoFlux|.  The user can change the global {\tt  fracCDM2} parameter before the calculation of these observables
to take into account the fact that the value of the dark matter fraction in the Milky Way could be different than   in the early Universe.

The \verb|darkOmega| function  is the same as in previous versions and is appropriate for models with only one dark matter candidate since it does not distinguish the  classes of the discrete  symmetry group. For example it will assume that all  particles whose name starts with one or two tildes belong to the same dark sector and are in thermal equilibrium. This is in general not the case if the discrete symmetry distinguishes two dark matter sectors. The \verb|darkOmega| function  should therefore be used only for models with one dark matter sector. 

The DM nucleon amplitude and cross section relevant for direct detection is computed for each DM candidate with the help of the 
 routine \\  
 $\bullet$ \verb|nucleonAmplitudes( CDM, qBOX,pAsi,pAsd,nAsi,nAsd)|\\
where the first parameter is the name of DM particles. All other parameters have the same meaning as in  previous versions.
Here there is no rescaling to account for the dark matter fraction of each component.

The \micro~ routines for  model independent analyses  contained in the \verb|mdlIndep| directory do not take into account the  possibility of two DM particles. These routines depend only on the global parameter {\tt Mcdm}.
All facilities of previous versions of  \micro~\cite{Belanger:2013oya} are included in this version except for the option to compute the  relic density and DM observables when there is an initial  ${\rm DM}  - \overline{{\rm DM}}$ asymmetry.   
A complete list of \micro~ routines is provided in the manual contained in  the \verb|man| directory.

\section{Example} 
\label{sec:example}

\begin{figure}[htb]
\label{fig:y}
\centering
\includegraphics[width=13cm]{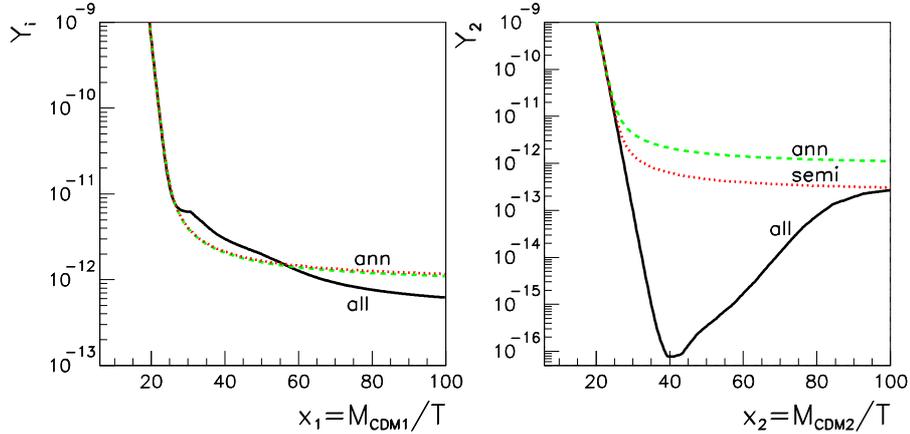}
\vspace{-.8cm}
\caption{Abundance ($Y_i$) as a function of $x=M_{\rm CDM1}/T$ for each DM particle in the doublet and singlet $Z_4$ model  when including all channels (full), only annihilation channels (dash), and also semi-annihilation channels (dot). Note that for CDM1 (left plot), adding semi-annihilation channels induces only a few percent variation in the abundance.
}
\end{figure}

As an example we will consider an extension of the SM with  two Higgs doublets and a singlet and a discrete $Z_4$ symmetry ~\cite{Belanger:2014bga}. The 
potential of the model is given in Eqs.~(\ref{eq:Vc}, \ref{eq:pot:Z4}), the independent parameters are chosen as the
masses of the scalars
$M_h, M_H, M_S, M_{H^0},M_{A^0}$ and 8 of the $\lambda_i$'s, see Table~\ref{tab:Z4}. The first dark sector contains only $S$ while the dark sector 2 contains the doublet $H_0, A^0, H^\pm$, with either $H^0$ or $A^0$ as the possible dark matter. For the set of input parameters defined in Table~\ref{tab:Z4}, 
The evolution of the abundance for the two dark matter candidates $S, H$ 
is illustrated in Figure~\ref{fig:y}. Furthermore the abundances are  compared with the case where DM conversion and/or semi-annihilation is ignored.

\begin{table}[htbp]
 \caption{Input parameters of $Z_4$ model}
 \label{tab:Z4}
\begin{center}
\begin{tabular}{|l |l|l||l|l|l||l|l|l|}
\hline
Name & Value       &         & Name& Value &  & Name & Value       &                                 \\  \hline
  \verb|Mh|      & 125.89 & $m_h$&     \verb|la2|&0.8162&$\lambda_2$& \verb|laS1|  &0.2367  &$\lambda_{S_{1}}$\\
 \verb|Msc|  & 578.0& $m_s$ & \verb|la3|&-0.1723& $\lambda_3$       &    \verb|laS2|  &-1.139  &$\lambda_{S_{2}}$ \\
 \verb|MHX|    &  895.5       & $m_H$&  \verb|laS|  &1.9121  &$\lambda_{S}$ & \verb|laS12|  &0.7629  &$\lambda_{S_{12}}$\\
 \verb|MH3|     &  900.6          & $m_A$  &\verb|lapS|  &1.017  &$\lambda'_{S}$ &\verb|laS21|  &-0.2054  &$\lambda_{S_{21}}$ 
\\
  \verb|MHC|      &  895.64           & $m_{H+}$ &&&&&&\\
\hline
\end{tabular}
\end{center}
\end{table}

\noindent
Running \micro~   for the benchmark point of Table~\ref{tab:Z4} (benchn3.par) with the options \verb|MASSES_INFO,OMEGA|,
\verb|INDIRECT_DETECTION,DIRECT_DETECTION|
and the switches \verb|VWDECAY=1|,\verb|VZDECAY=1| will lead to the following output

\begin{verbatim}
Dark matter candidate is '~sc' with spin=0/2 mass=5.78E+02
Dark matter candidate is '~~X' with spin=0/2 mass=8.96E+02

=== MASSES OF HIGG AND ODD PARTICLES: ===
Higgs masses and widths
      h   125.89 4.26E-03

Masses of odd sector Particles:
~sc  : Msc   =   578.0 || ~~X  : MHX   =   895.5 || ~~H+ : MHC   =   895.6 
~~H3 : MH3   =   900.6 || 

==== Calculation of relic density =====
omega1=5.91E-02
omega2=6.14E-02

==== Indirect detection =======
    Channel          vcs[cm^3/s]
==================================
 annihilation cross section 5.89E-26 cm^3/s
 contribution of processes
 annihilation cross section 5.89E-26 cm^3/s
 contribution of processes
  ~sc,~~X -> Z ~Sc      1.83E-01
  ~Sc,~~X -> Z ~sc      1.83E-01
  ~~X,~~X -> W+ W-      7.47E-02
  ~sc,~sc -> W- ~~H+    7.24E-02
  ~Sc,~Sc -> W+ ~~H-    7.24E-02
  ~sc,~~X -> h ~Sc      6.15E-02
  ~Sc,~~X -> h ~sc      6.15E-02
  ~~X,~~X -> Z Z        5.20E-02
  ~sc,~Sc -> W+ W-      3.95E-02
  ~sc,~sc -> Z ~~X      2.82E-02
  ~Sc,~Sc -> Z ~~X      2.82E-02
  ~sc,~sc -> h ~~H3     2.30E-02
  ~Sc,~Sc -> h ~~H3     2.30E-02
  ~sc,~Sc -> Z Z        1.96E-02
  ~sc,~Sc -> h h        1.91E-02
  ~~X,~~X -> h h        1.09E-02
  ~sc,~sc -> Z ~~H3     8.50E-03
  ~Sc,~Sc -> Z ~~H3     8.50E-03
  ~sc,~sc -> h ~~X      7.27E-03
  ~Sc,~Sc -> h ~~X      7.27E-03
  ~sc,~sc -> W+ ~~H-    5.57E-03
  ~Sc,~Sc -> W- ~~H+    5.57E-03
  ~sc,~Sc -> t T        3.85E-03
  ~~X,~~X -> t T        8.32E-04
sigmav=5.89E-26[cm^3/s]
Photon flux  for angle of sight f=0.10[rad]
and spherical region described by cone with angle 0.10[rad]
Photon flux = 2.13E-16[cm^2 s GeV]^{-1} for E=289.0[GeV]
Positron flux  =  1.01E-14[cm^2 sr s GeV]^{-1} for E=289.0[GeV] 
Antiproton flux  =  1.75E-13[cm^2 sr s GeV]^{-1} for E=289.0[GeV] 

==== Calculation of CDM-nucleons amplitudes  =====
CDM[antiCDM]-nucleon micrOMEGAs amplitudes for ~sc 
proton:  SI  1.813E-09 [1.813E-09]  SD  0.000E+00 [0.000E+00]
neutron: SI  1.831E-09 [1.831E-09]  SD  0.000E+00 [0.000E+00]
CDM[antiCDM]-nucleon cross sections[pb]:
 proton  SI 1.433E-09 [1.433E-09] SD 0.000E+00 [0.000E+00]
 neutron SI 1.461E-09 [1.461E-09] SD 0.000E+00 [0.000E+00]
CDM[antiCDM]-nucleon micrOMEGAs amplitudes for ~~X 
proton:  SI  -8.929E-10 [-8.929E-10]  SD  0.000E+00 [0.000E+00]
neutron: SI  -9.017E-10 [-9.017E-10]  SD  0.000E+00 [0.000E+00]
CDM[antiCDM]-nucleon cross sections[pb]:
 proton  SI 3.473E-10 [3.473E-10] SD 0.000E+00 [0.000E+00]
 neutron SI 3.543E-10 [3.543E-10] SD 0.000E+00 [0.000E+00]
\end{verbatim}

\section*{Acknowledgements}

We thank Kristjan Kannike and Marrti Raidal for their collaboration on two dark matter models  that lead to the development of this code. 
This work was  supported in part by the LIA-TCAP of CNRS,  by the French ANR, Project DMAstro-LHC, ANR-12-BS05-0006,
and by the {\it Investissements d'avenir}, Labex ENIGMASS.
The work of AP  was  also supported by the Russian foundation for Basic Research, grant
RFBR-12-02-93108-CNRSL-a.

\appendix
\section*{Appendix}

As another example of a two-component DM model, we consider a model that contains two singlets  $S_1$  and $S_2$ in addition to the SM Higgs doublet, $H$. We impose  a  $Z_5$ discrete symmetry
with $X_{H}=0$ , $X_{S_1}=1/5$ and $X_{S_2}=-3/5$. The potential reads

\begin{equation}
\begin{split}   
V_{Z_{5}} &= V_{0} 
          + \frac{\lambda_{31}}{2} (S_1^{3} S_2 + S_1^{\dagger 3}S_{2}^{\dagger}) +\frac{\lambda_{32}}{2} (S_2^{3} S_{1}^{\dagger}
          + S_2^{\dagger 3} S_{1})\\
          &+ \frac{\mu_{SS1}}{2} (S_1^{2} S_2^\dagger + S_1^{\dagger 2}S_{2})
          + \frac{\mu_{SS2}}{2} (S_1 S_2^2 + S_1^{\dagger}S_{2}^{\dagger 2}),
\end{split}
\label{eq:pot:Z5}
\end{equation}

where 

\begin{equation}
\begin{split}   
  V_{0}&=  \mu_{1}^{2} |{H}^{2}| + \lambda_{1} |{H}^{4}|
   + \mu_{S1}^{2} |S_1|^{2}+ \mu_{S2}^{2} |S_2|^{2}
+ \lambda_{41} |S_1|^{4}+ \lambda_{42} |S_2|^{4} + \lambda_{412} |S_1|^{2} |S_2|^{2} \\
  &+ \lambda_{S1} |S_1|^{2} |H|^{2}+ \lambda_{S2} |S_2|^{2} |H|^{2}
  \end{split}
\label{eq:Vc5}
\end{equation}
is the part of the potential valid for any choice of discrete symmetry $Z_N$.
The input parameters of the model are the masses  $M_h$,$M_{S1}$, $M_{S2}$ and the couplings $\lambda_{41},\lambda_{42},\lambda_{412}$, $\lambda_{31},\lambda_{32}$, $\lambda_{S1},\lambda_{S2}$, $\mu_{SS1},\mu_{SS2}$. This model features only one particle in each of the dark sector and  includes interactions between the two dark sectors that were not present in the $Z_4$ model considered in section~\ref{sec:example}, for example $S_1 S_1\rightarrow S_1^\dagger S_2^\dagger$. 
This model is provided in the directory {\tt Z5M} together with the {\tt Lanhep} source code to create the appropriate {\tt CalcHEP} model files.

\providecommand{\href}[2]{#2}\begingroup\raggedright\endgroup

\end{document}